\documentclass{article}

\usepackage[preprint]{neurips_2025}
\workshoptitle{AI for Music}

\usepackage[utf8]{inputenc} 
\usepackage[T1]{fontenc}    
\usepackage{hyperref}       
\usepackage{url}            
\usepackage{booktabs}       
\usepackage{amsfonts}       
\usepackage{nicefrac}       
\usepackage{microtype}      
\usepackage{xcolor}         
\usepackage{tabularx}
\usepackage{multirow}

\usepackage{amsmath}
\usepackage{graphicx}
\usepackage[colorinlistoftodos]{todonotes}
\usepackage{tabularx}
\usepackage{float}
\usepackage{booktabs}
\usepackage{pdfpages}
\usepackage{multirow}
\usepackage{geometry}
\usepackage{longtable}
\usepackage{hyperref}

\usepackage{natbib}

\title{CoComposer: LLM Multi-agent Collaborative  Music Composition}

\author{%
  Peiwen Xing, Aske Plaat, Niki van Stein \\
  LIACS, Leiden University, Netherlands \\
  \texttt{aske.plaat@gmail.com} \\
}

\begin{document}

\maketitle

\begin{abstract}
Existing AI Music composition tools  are limited in
generation duration,  musical quality,  and  controllability. We introduce CoComposer, a multi-agent system that 
consists of five collaborating  agents, each with a  task based on the traditional music composition workflow. 
Using the AudioBox-Aesthetics system,  we experimentally evaluate CoComposer on four compositional criteria. We test with three LLMs (GPT-4o, DeepSeek-V3-0324, Gemini-2.5-Flash), and find (1) that CoComposer outperforms  existing multi-agent LLM-based systems in  music quality, and (2) compared to a single-agent system, in production complexity. Compared to  non-LLM  MusicLM, CoComposer has better interpretability and editability, although MusicLM still produces better music.
\end{abstract}

\section{Introduction}

%

The qualities of large language models (LLMs) \citep{zhao2023survey} for music composition are actively explored, although 
%
current  systems still have severe limitations. First, the  duration of generated pieces is limited. For example, Google's MusicFX \citep{musicfx} can only output 30-second music. 
Second, mainstream platforms such as Suno \citep{suhailudheensuno} do not support the input of long prompt words. Third, for professional musicians the generated music  has a rather "toy-like" nature. Fourth, the controllability is poor: there are deficiencies in the detailed editing of music, and it is difficult to accurately achieve the refined adjustments and creative intentions of professionals for music. 
%
To address the above challenges, we have created a multi-agent symbolic music composition system, CoComposer. 
%
Our main contributions are as follows:
(1) We introduce a multi-agent collaborative composition system where agent roles closely follow the traditional music composition. The system uses generic LLMs.
(2) In order to evaluate CoComposer objectively, we use  AudioBox-Aesthetics, which shows that we outperform other state of the art agentic-LLM composition systems. 
(3) We open source our system, and  provide directions for further improvement.\footnote{\url{https://github.com/PhotonCombiner/CoComposer}}

\paragraph{Related Work}

The fields of audio and music creation are witnessing a  technological revolution \citep{briot2020deep,agostinelli2023musiclm,copet2023simple,yuan2024chatmusician,fugatto,liu2024mumu,lam2025analyzable,deng2024composerx,yan2024measure}. 
%
The core capabilities of LLM lie in understanding the semantics, grammar, and context logic of natural language \citep{minaee2024large}. LLMs are also known to understand affective reasoning \citep{broekens2023fine}, and agentic LLMs allow collaboration, tool use, and role specialization \citep{plaat2025agentic}. 

Agents in a multi-agent system have autonomous decision-making capabilities \citep{guo2024large,wong2021deep}. Complex task objectives are achieved through  interaction and collaboration, 
making it  suitable for complex scenarios such as polyphonic music composition tasks.
%
%
A popular open-source natural languge framework is AutoGen \citep{wu2024autogen}, which can simplify the communication between music-composition agents. 
%
Role-playing language agents simulate the language style, knowledge background, and behavioral characteristics of a specific role, to achieve interaction that conform to the agent's role  \cite{chen2024persona}. 
ComposerX is a system in which the GPT-4-turbo model plays the role of a professional composer  \citep{deng2024composerx}.  ComposerX consists of a leader-agent, a melody-agent, a harmony-agent, an instrument-agent, a reviewer-agent, and an arrangement-agent. It receives simple instructions in  user prompts, and generates ABC notation \citep{walshaw2021abc}, controlled by AutoGen.  
In an evaluaiton with humans experts, the quality of the generated works was found to be comparable to that of dedicated symbolic music generation system that require substantial computational resources and data support.
%
These LLM-based system are prompt-based: all learning takes place in-context \citep{brown2020language}, without changing the parameters of the neural network.
In contrast,  NotaGen \citep{wang2025notagen} is a multi-agent system that adopts a more elaborate training scheme, with pretraining and finetuning the network parameters. 
In our approach, we use the simpler in-context learning approach.


Music composition, as a complex and creative art activity, encompasses several fundamental elements. 
Through the creative combination of melody, harmony, rhythm, and larger structures, a musical piece is composed.
%
The general process of polyphonic music composition follows the logic of {\em main melody - arrangement and orchestration - revision}. First, we determine the main melody, clarifying its tonality, rhythmic pattern, and emotional tone, serving as the basic skeleton of the polyphonic texture. Subsequently, we design the accompaniment part, which needs to form a contrapuntal relationship with the main melody.
A multi-agent approach allows these elements to be performed by specialized agents.
%
Through multiple rounds of iteration, a complete work is formed. 

In this work we explore the possibility of multi-agent architectures for polyphonic and multi-track multi-role symbolic music generation with hierarchical structures. We focus on generation of ABC notation, using a midi backend for sound generation.


\section{Method}
Based on the related work, we use an agentic LLM system, inspired by the limitations of ComposerX/AutoGen (see appendix), with  agent roles that more closely collaborate and follow traditional music composition roles. Our system is called CoComposer, for Collaborative Composer.  We perform a less-subjective performance evaluation with AudioBox-Aesthetics.

The CoComposer system employs five role-playing language agents: Leader Agent, Melody Agent, Accompaniment Agent, Revision Agent, Review Agent, see Table \ref{tab:cocomposer}. AutoGen enables collaborative cooperation, forming a compact architecture of "task decomposition - creative execution - corrective feedback". The agents organize ``group chats'' where each agent takes turns to speak in sequence. 
The creative process of each agent in the system is divided into two phases.

\begin{table}
    \centering
    {\footnotesize
    \begin{tabularx}{\textwidth}{|>{\raggedright}p{4cm}|X|}\hline
         Agent Name& Core Responsibility\\\hline
         Leader Agent& Acts as the team leader, receives and parses user requirements, decomposes tasks and assigns them to other agents.\\\hline
         Melody Agent& Serves as one of the core creative entities, responsible for main melody composition, selecting appropriate musical instruments, and outputting the required ABC notation content.\\\hline
         Accompaniment Agent& 	Serves as one of the core creative entities, responsible for accompaniment composition, selecting appropriate musical instruments, and outputting the required ABC notation content.\\\hline
         Revision Agent& 	Focuses on ABC notation revision, detects and fixes format and rhythm errors in the notation.\\\hline
         Review Agent& Reviews works from the perspective of music theory and provides optimization suggestions.\\ \hline
    \end{tabularx}
    }
    \caption{Names and main responsibilities of agents in the CoComposer system }
    \label{tab:cocomposer}
\end{table}


{\em Initialization Creation Phase}
1) The Leader Agent first analyzes the instruction prompt. It assigns the main melody creation task to the Melody Agent and the accompaniment creation task to the Accompaniment Agent. 
2) The Melody Agent independently completes the creation of the main melody in ABC format, attaches MIDI instrument information, and then uploads the results to the shared dialogue pool. 
3) The Accompaniment Agent designs the accompaniment based on the melody, attaches MIDI instrument information, and uploads the accompaniment content to the dialogue pool. 
4) The Revision Agent checks the notated content. It corrects timing errors and format errors without changing creative content, and then sends the corrected content back.
5) The Review Agent extracts the corrected content, reviews it from five dimensions (such as melodic structure and harmony) and formulates specific improvement suggestions.

{\em Iterative Creation Phase} 
Next, in the iterative phase, the Leader Agent checks the review suggestions of the Review Agent in the dialogue pool, and determines the key parts that need to be optimized, instructing the Melody Agent and the Accompaniment Agent for modifications, after which the Revision Agent conducts another check. Etc.


The design of the prompt words revolves around the functional positioning and collaboration logic of the roles, aiming to ensure that each Agent has a clear division of labor and efficient cooperation. The specific prompts can be found in the appendix.
Compared to ComposerX, the core improvements of CoComposer are a better  agent structure and role division, and optimized prompts. 
CoComposer only needs five agents to complete the same composition task. The reduction in the number of agents is not simply a decrease, but is based on the typical composition process of multi-voice music. In real  compositions, usually one person is responsible for writing the main melody, another person designs the accompaniment, and then they make modifications based on the guidance of a third party.
Compared with a separate Instrument Agent, our design reduces the number of communication rounds between agents, making the system more efficient.

\section{Evaluation}
In our evaluation, we compared CoComposer to ComposerX and to MusicFX \citep{musicfx,agostinelli2023musiclm}.
CoComposer used GPT-4o, DeepSeek-V3-0324, and Gemini-2.5-Flash as LLMs respectively, and ComposerX used GPT-4o.

The total number of user input prompts used in this experiment is 20. These prompts are all from the prompt set constructed by ComposerX, among which 10 have been specifically abridged, 
to leave more freedom for system creation. The prompt set covers various music styles and instrument types, providing standardized input support for system verification and comparative experiments. The complete prompt set can be found in the appendix. An example prompt is: {\em 
  Retro Video Game Adventure: Develop a playful chiptune piece in F major with a fast tempo. The chord progression should be F, G, Am, Bb, spanning 32 bars. Use 8-bit synth and electronic drums. The 8-bit synth should provide nostalgic, catchy melodies reminiscent of classic video games, while the electronic drums should add a rhythmic, upbeat backing. This track should evoke the excitement and adventure of retro video gaming.}


We used Meta's AudioBox-Aesthetics \citep{tjandra2025meta} score prediction model for automated  evaluation. 
AudioBox is based on the Transformer architecture and evaluates music aesthetics from the following four dimensions. 
(1) Production Quality (PQ) focuses on the technical aspects including clarity, dynamics, frequencies, and spatialization of the audio; (2) Production Complexity (PC) focuses on the complexity of an audio scene measured by the number of audio components; (3) Content Enjoyment (CE) focuses on the subjective quality of an audio piece including emotional impact, artistic skill, artistic expression, as well as subjective experience; (4) Content Usefulness (CU) focuses on leveraging the audio as source material for content creation.







The evaluation results of CoComposer are shown in see Table \ref{tab:results}. Experiment 1 compares CoComposer against ComposerX. Experiment 2 compares the multi-agent composers against a single agent. Experiment 3 compares 3 different LLMs. Experiment 4 compares CoComposer against MusicFX.


\begin{table}
    \centering
    {\footnotesize
    \begin{tabularx}{\textwidth}{p{1cm}p{4.45cm}>{\centering\arraybackslash}p{1cm}>{\centering\arraybackslash}p{1cm}>{\centering\arraybackslash}p{1cm}>{\centering\arraybackslash}p{1cm}p{2cm}}
    \toprule
                                &\textbf{Music Composition Systems} &\textbf{CE}   &\textbf{CU}   &\textbf{PC}   &\textbf{PQ}   &\textbf{Gen. Success}\\
    \midrule
    \multirow{2}{=}{Exper 1} &CoComposer with GPT-4o           &\textbf{6.75} &\textbf{7.76} &\textbf{4.13} &\textbf{7.86} &100\%\\
                                  &ComposerX with GPT-4o            &6.52          &7.61          &3.72          &7.76          &100\%\\
    \hline
    \multirow{3}{=}{Exper 2} &CoComposer with GPT-4o           &\textbf{6.75} &7.76          &\textbf{4.13} &7.86          &100\%\\
                                  &ComposerX with GPT-4o            &6.52          &7.61          &3.72          &7.76          &100\%\\
                                  &single-agent with GPT-4o         &6.72          &\textbf{7.77} &3.92          &\textbf{7.88} &100\%\\
    \hline
    \multirow{3}{=}{Exper 3} &CoComposer  Deepseek-V3-0324 &\textbf{6.77} &7.70          &3.98          &7.85          &100\%\\
                                  &CoComposer Gemini-2.5-Flash &6.37          &7.57          &3.92          &7.73          &100\%\\
                                  &CoComposer  GPT-4o           &6.75          &\textbf{7.76} &\textbf{4.13} &\textbf{7.86} &100\%\\
    \hline
    \multirow{4}{=}{Exper 4} &CoComposer Deepseek-V3-0324 &6.77          &7.70          &3.98          &7.85          &100\%\\
                                  &CoComposer  Gemini-2.5-Flash &6.37          &7.57          &3.92          &7.73          &100\%\\
                                  &CoComposer  GPT-4o           &6.75          &7.76          &4.13          &\textbf{7.86} &100\%\\
                                  &MusicFX                          &\textbf{7.37} &\textbf{7.93} &\textbf{4.96} &7.84          &\\
    \bottomrule
    \end{tabularx}
    }
    \caption{Summary table of the results of four experiments}
    \label{tab:results}
\end{table}

The generation success rate of all systems (except MusicFX) is 100\%, indicating that the LLM-based  systems generate symbolic music reliably. 
CoComposer is  superior to ComposerX in subjective aesthetic experience (CE, CU), creative complexity (PC), and production quality (PQ), suggesting that the system's streamlining of the agent architecture (replacing 6 agents with 5), the {\em creation-orchestration synchronization} process, and the prompt optimization effectively improve the quality of music creation.

Compared to the single-agent system, CoComposer has a significant advantage in production complexity (PC) and is far superior to ComposerX.
Experiment 3 compares the performance of CoComposer for  different LLMs (GPT-4o, DeepSeek-V3-0324, Gemini-2.5-Flash). GPT-4o performs best overall.
Experiment 4 compares CoComposer (three types of LLMs) with the dedicated music generation model MusicFX. The results show that MusicFX significantly leads in terms of subjective experience and complexity.
Although CoComposer is inferior to MusicFX in aesthetic indicators, it has  advantages in "interpretability" and "editability", since it is open and uses the ABC notation.

\section{Conclusion}
In summary, we find that the use of a multi-agent approach can improve the music quality created by an LLM. Experiments show that through reasonable agent division of labor design and high-quality prompts, CoComposer has achieved an improvement in production complexity compared to the single-agent system, is  on a par in terms of content enjoyment and content usefulness, and has an overall better performance.
CoComposer 
uses natural language as the input interface: users do not need to master musical theory knowledge or professional tools, which greatly reduces the technical barriers to music creation. At the same time, it is built based on existing LLMs and open-source frameworks. There is no need for additional large-scale music data pre-training,
significantly reducing the system's research, development, and deployment costs.
The system adopts the ABC notation as an intermediate carrier.
Users can directly view and edit the notation content, and understand the composition logic of melodies and harmonies.





Although CoComposer is close to MusicFX in Production quality (PQ) and has the advantage of interpretability, it still significantly lags behind the dedicated music model in core aesthetic dimensions (CE, CU, PC). This result is in line with the difference in technical positioning: MusicFX is a dedicated model optimized for music generation, while CoComposer is a multi-agent system based on a general-purpose LLM. Its core value lies in the {\em low-threshold, high-controllability} creative process, rather than competing with the dedicated model in absolute aesthetic quality.

\paragraph{Future Work}


The ABC notation relies on the MIDI standard sound library for audio conversion, and the core of the MIDI sound library is to simulate traditional acoustic instruments (such as pianos, violins, drum sets, etc.). It is unable to define the sounds in modern music that rely heavily on computer synthesis.
The system also relies on the native capabilities of general LLMs without special fine-tuning for music composition tasks. This leads to limitations in the model's understanding and generation capabilities for complex musical structures (such as polyphonic counterpoint and large-scale musical forms).
Although the system can generate music works that conform to style characteristics based on text descriptions, there are  boundaries at the level of subjective creativity. As a carrier of human emotional expression, music often contains the personal experiences, cultural backgrounds, and instant inspirations of the creator.
%
Finally, we suggest to design a special feedback analysis agent to transform users' fragmented feedback into structured creative instructions (for example, mapping "this melody is too cheerful" to "reduce the tempo by 10 BPM and add minor-key color"). At the same time, introduce a "memory mechanism" to record users' preference patterns in multiple rounds of creation (such as preferring string timbres and tending to compact rhythm patterns), gradually optimize the generation direction, and achieve a personalized creative experience.




\bibliographystyle{plainnat}
\bibliography{cocomposer}

\clearpage

\appendix

\section*{Appendix}

\section{CoComposer Prompts}

The prompt words for the Leader Agent are shown in Table \ref{tab:leader}. The core responsibility of the Leader Agent is demand transformation and task assignment. The prompt words clearly require it to extract key musical elements (such as style, tonality, musical instruments, etc.) from the customer's requirements, and disassemble the tasks and assign them to the Melody Agent and the Accompaniment Agent.

\begin{table}[H]
    \centering
    {\footnotesize
    \begin{tabularx}{\textwidth}{|X|}
        \hline
        You are the leader of a music production team, which includes Melody Agent, Accompaniment Agent, Revision Agent and Reviewer Agent.\\
        You will receive the request from the client, which will be a breif desciption of the kind of music they want.\\
        You need to carefully analyze the musical elements given in the request, which usually includes the title, genre, key, chord pregression, instruments, tempo, rhythm of the music.\\
        After examing the client's request, you are responsible for decomposing it into subtasks, and assign the subtasks to only Melody Agent and Accompaniment Agent in your team.\\
        \hline
    \end{tabularx}
    }
    \caption{Leader Agent prompt}
    \label{tab:leader}
\end{table}

The prompt words for the Melody Agent are shown in Table \ref{tab:melody}. The Melody Agent focuses on the creation of the main melody. The prompt words limit it to generate a single melody line in the ABC Notations format and mark the MIDI instrument information. At the same time, it is required to optimize the work according to the feedback of the Reviewer Agent. This design not only ensures the professionalism of melody creation but also provides a unified format basis for subsequent collaboration. In addition, it is clearly required to "not output other words". (The Accompaniment Agent and the Revision Agent also have this requirement). This design ensures that the output content only contains the musical score information, avoiding the interference of redundant information on the process, reducing the system operation cost and improving the efficiency of cross-agent collaboration.

\begin{table}[H]
    \centering
    {\footnotesize
    \begin{tabularx}{\textwidth}{|X|}
        \hline
        You are skillful musician, especially in melody.\\
        You will compose a single-line melody based on the client's request and assigned tasks from the Leader.\\
        You will decide appropriate instrument for your melody and annotate them using \%\%MIDI program.\\
        You must output your work in ABC Notations.\\
        Example:\\
        ```\\
        X:1\\
        T:Journey Through the Highlands\\
        M:6/8\\
        L:1/8\\
        Q:3/8=100\\
        K:A\\
        V:1 name="Bagpipe Lead" \%\%MIDI program 109\\
        |: "A"A2e c2A | "D"dcd B2G | "E"EFE G2B | "A"A2A A3 :|\\
        |: "A"ece a2f | "D"d2D F3 | "E"G2E B3 | "A"A2A A3 :|\\
        ```\\
        Markdown your work using ```    ``` to the client.\\
        After you receive the feedback from the Reviewer Agent, please improve your work according to the suggestions you were given.\\
        Note: Only output the sheet music in the specified ABC Notations format, with no other text.\\
        \hline
    \end{tabularx}
    }
    \caption{Melody Agent prompt}
    \label{tab:melody}
\end{table}

The prompt words for the Accompaniment Agent are shown in Table \ref{tab:accom}. The core of the Accompaniment Agent is to create an accompaniment that complements the melody. The prompt words emphasize that it needs to design the harmonic progression, accompaniment texture, and instrument combination based on the melody, and clearly distinguish the melody and accompaniment parts in the ABC Notations, ensuring the coordination of the two in rhythm and tonality and avoiding conflicts between the accompaniment and the melody.

\begin{table}[H]
    \centering
    {\footnotesize
    \begin{tabularx}{\textwidth}{|X|}
        \hline
        You are a skilled musician specializing in accompaniment composition, particularly in designing accompaniment patterns, harmonic support, and instrumental texture for melodies.\\
        Your core task is to create a complementary accompaniment for the melody provided by the Melody Agent. You will compose the accompaniment based on the client's request and assigned tasks from the Leader. This includes:\\
        1. Designing harmonic progressions that support the melody's tonality and emotion.\\
        2. Creating appropriate accompaniment textures (e.g., arpeggios, block chords, rhythmic patterns) that enhance the melody without overshadowing it.\\
        3. Selecting suitable instruments for the accompaniment (e.g., piano, bass, strings) that match the style and complement the melody's lead instrument.\\
        Ensure the accompaniment aligns with the melody in rhythm, key, and structure, and maintains a balanced relationship (supporting rather than competing).\\
        You must output your work in ABC Notations, clearly distinguishing between the melody (from Melody Agent) and your accompaniment parts.\\
        Example:\\
        ```\\
        X:1
        T:Journey Through the Highlands\\
        M:6/8\\
        L:1/8\\
        Q:3/8=100\\
        K:A\\
        V:1 name="Bagpipe Lead" \%\%MIDI program 109\\
        |: "A"A2e c2A | "D"dcd B2G | "E"EFE G2B | "A"A2A A3 :|\\
        |: "A"ece a2f | "D"d2D F3 | "E"G2E B3 | "A"A2A A3 :|\\
        V:2 name="String Harmony" \%\%MIDI program 48\\
        |: E2c A2F | F2D D2E | G2E F2D | E2E E3 :|\\
        |: c2B A2c | B2A G2F | A2F E2D | C2C C3 :|\\
        ```\\
        Markdown your work using ```    ``` to the client.\\
        After you receive the feedback from the Reviewer Agent, please imporove your work according to the suggestions you were given.\\
        Note: Only output the sheet music in the specified ABC Notations format, with no other text.\\
        \hline
    \end{tabularx}
    }
    \caption{Accompaniment Agent prompt}
    \label{tab:accom}
\end{table}

The prompt words for the Revision Agent are shown in Table \ref{tab:rev}. The Revision Agent focuses on the revision of the ABC notation. The prompt words strictly limit it to only correct timing errors (such as the inconsistency between the measure duration and the time signature) and format errors (such as violations of the ABC specifications), and follow the "minimum intervention principle" without changing the creative content to ensure the integrity of the original creative intention.

\begin{table}[H]
    \centering
    {\footnotesize
    \begin{tabularx}{\textwidth}{|X|}
        \hline
        You are an expert in ABC Notations, specializing in error correction.\\
        You will receive an ABC notation draft with potential timing errors and must follow this core principle: ONLY modify parts with confirmed errors; leave all correct content unchanged.\\
        Your tasks are strictly limited to:\\
        1. Identify timing errors where a measure's total duration does not match the specified time signature (M:). For these cases only, adjust note durations (e.g., extend/shorten notes, split/merge notes, or add rests) to fix the mismatch. \\
        2. Fix formatting errors (e.g., incorrect voice labels, missing \%\%MIDI programs, or malformed chord symbols) only where they clearly violate ABC notation standards.
        Markdown your work using ```    ``` to the client.\\
        Note: Only output the sheet music in the specified ABC Notations format, with no other text.\\
        """\\
        \hline
    \end{tabularx}
    }
    \caption{Revision Agent prompt}
    \label{tab:rev}
\end{table}

The prompt words for the Review Agent are shown in Table \ref{tab:review}. The Review Agent undertakes the functions of comprehensive review and optimization guidance. The prompt words require it to conduct a strict evaluation from five dimensions: melodic structure, harmonic counterpoint, rhythmic complexity, orchestration timbre, and overall form, provide specific improvement suggestions for the performance of each Agent, and promote the iterative upgrade of the work.

\begin{table}[H]
    \centering
    {\footnotesize
    \begin{tabularx}{\textwidth}{|X|}
        \hline
        You are a skillful musician, you are expertized in music theory.\\
        You need to be very strict and critical about their work.\\
        You will check the entire work and provide constructive critics.\\
        You will critize on each agent's performance so they can improve the quality of their work.\\
        The agents are accessed based on:\\
        Melodic Structure: Assess the flow, thematic development, and variety in pitch and rhythm.\\
        Harmony and Counterpoint: Check how harmonies support the melody, effectiveness of counterpoint, and chord progressions.\\
        Rhythmic Complexity: Evaluate the rhythm's contribution to interest, its interaction with the melody, and dynamic changes.\\
        Instrumentation and Timbre: Look at the appropriateness of instruments, blending of timbres, and use of dynamics.\\
        Form and Structure: Analyze the overall structure, transitions, and how sections are connected and concluded.\\
        \hline
    \end{tabularx}
    }
    \caption{Review Agent prompt}
    \label{tab:review}
\end{table}

\section{Evaluation Prompts}

We have organized the 20 user input prompt words used in the experiments (see Table \ref{tab:eval}).

\footnotesize{
\begin{longtable}{|c|p{0.91\textwidth}|}
\caption{Twenty user input prompts}\label{tab:eval} \\
\hline
 & \textbf{User Input Prompts} \\
\hline
\endfirsthead

\hline
 & \textbf{User Input Prompts} \\
\hline
\endhead

\endfoot

\endlastfoot
         1& Vintage French Chanson: Create a nostalgic chanson piece in C major with a slow tempo. The chord progression will be C, Am, Dm, G, played over 16 bars. Utilize accordion, violin, and upright bass. The accordion should lead with its melodious and expressive sound, the violin should add a romantic and wistful quality, and the upright bass should provide a warm, supporting foundation. This composition should evoke the charm and sentimentality of a vintage French chanson.\\\hline
         2& River Journey: Develop a composition that follows the flow of a river, using a motif of fluid, meandering melodies and a progression like C-G-Am-Em. Incorporate sounds that mimic the gurgling water and wildlife along the riverbank, set to a tempo that's both tranquil and lively.\\\hline
         3& Romantic Parisian Cafe: Create a romantic French piece in F major, following the chord progression F-Bb-C7-F. Use instruments like the accordion, violin, and gypsy guitar to set a moderate and romantic tempo. The rhythm should be sensual, much like the ambiance of a cafe in Paris, capturing the romantic and charming essence of the city of love.\\\hline
         4& Journey Through the Highlands: Compose a piece that reflects the rugged beauty of the Scottish Highlands. Use a bagpipe lead with a chord progression of A-D-E-A. The tempo should be brisk, with a rhythm that feels like a spirited walk through rolling hills and green valleys. Incorporate the sounds of nature to enhance the outdoor ambiance.\\\hline
         5& Tropical Rainforest Rhapsody: Compose a piece inspired by the lushness of a tropical rainforest, using a motif of intricate, layered rhythms and a progression like Dm-C-Bb-A. Include sounds that mimic raindrops and wildlife, creating a rich tapestry of natural harmony, set to a medium tempo.\\\hline
         6& Urban Jazz Nights: Write a jazz piece that captures the essence of a vibrant city at night, using a motif of smooth, flowing lines and a progression like Fm7-Bb7-Ebmaj7-Abmaj7. Use a saxophone to lead the melody, set to a medium swing tempo, reflecting the rhythmic pulse of city nightlife.\\\hline
         7& Blues Alley Tale: Write a classic 12-bar blues in the key of E, using the standard I-IV-V progression (E7-A7-B7). Keep the tempo moderate to slow, allowing each chord to carry the emotional weight of the blues narrative. Add a soulful harmonica or guitar melody to enhance the storytelling aspect of the music.\\\hline
         8& Renaissance Faire Dance: Take a step back in time to a cheerful and historical Renaissance faire dance in G major. The chord progression should be G-C-D-G, highlighting instruments like the lute, recorder, and viola da gamba. Set a lively tempo and create a dance-like rhythm that captures the spirit of a Renaissance celebration. The emotion should be cheerful and historical.\\\hline
         9& Dreamy Indie Road Trip: Compose a dreamy indie pop piece in G major. The tempo should be medium, creating a relaxed and contemplative mood suitable for a road trip. The chord progression will be G, D, Em, C, and should unfold over 24 bars. The ensemble should include four voices, featuring acoustic guitar, synth, bass, and drums. The acoustic guitar should be the primary melodic driver, offering gentle, rhythmic strumming that evokes the feeling of a leisurely journey. The synth should add a layer of dreaminess, with ethereal pads or soft, melodic lines that enhance the song's contemplative nature. The bass should provide a solid, yet unobtrusive foundation, grounding the composition while allowing the other instruments to shine. Drums should maintain a steady, simple beat, echoing the steady pace of a road trip. This composition should encapsulate the essence of a dreamy, introspective journey, perfect for long drives along scenic routes.\\\hline
         10& Retro Video Game Adventure: Develop a playful chiptune piece in F major with a fast tempo. The chord progression should be F, G, Am, Bb, spanning 32 bars. Use 8-bit synth and electronic drums. The 8-bit synth should provide nostalgic, catchy melodies reminiscent of classic video games, while the electronic drums should add a rhythmic, upbeat backing. This track should evoke the excitement and adventure of retro video gaming.\\\hline
         11& Serenade Under the Moonlight: Craft a classical piece in the romantic style, conveying a melancholic mood. This composition should evoke the essence of a serene, moonlit night, filled with deep emotion and contemplation.\\\hline
         12& Summer Jazz Festival: Compose a lively and joyful jazz piece in the bebop style. This composition should evoke the lively atmosphere of a summer jazz festival.\\\hline
         13& Downtown Groove: Construct an energetic funk piece with a groovy style. This composition should embody the lively and vibrant atmosphere of a downtown scene, brimming with groove and energy.\\\hline
         14& Glacial Odyssey: Create a peaceful New Age composition with an ethereal style. This composition should transport the listener on a peaceful odyssey through pristine, icy realms, emphasizing the majesty and serenity of nature.\\\hline
         15& Neon City Lights: Craft a nostalgic synthwave piece with a retro style. This composition should transport the listener to a vibrant, neon-lit city, filled with the retro charm and wistful nostalgia of the synthwave genre.\\\hline
         16& Midnight Blues Café: Craft a soulful classic blues piece. The tempo should be slow, reflecting the introspective and emotional depth of the blues genre. This composition should evoke the ambiance of a dimly lit café at midnight, filled with the rich, heartfelt strains of classic blues.\\\hline
         17& Renaissance Fair Minuet: Create an elegant Baroque-style classical piece. The tempo should be moderato, reflecting the stately and dignified pace of a minuet. This composition should capture the grandeur and formality of a Renaissance fair, transporting the listener to a time of courtly dances and opulent celebrations, all encapsulated within the elegant framework of a Baroque minuet.\\\hline
         18& Arctic Electronica: Craft a cool, ambient techno piece in the electronic genre. The tempo should be moderate, capturing the crisp and sleek essence of Arctic-inspired sounds. This composition should encapsulate the feeling of a serene, icy environment, blending the chill of the Arctic with the warmth of electronic beats.\\\hline
         19& Reggae Beach Vibes: Develop a laid-back roots reggae piece. The tempo should be medium, embodying the relaxed and rhythmic nature of reggae music. This composition should transport the listener to a serene beach setting, where the rhythms of reggae blend seamlessly with the sounds of the ocean.\\\hline
         20& Urban Street Jazz: Compose a vibrant jazz fusion piece. The tempo should be fast, reflecting the dynamic and lively energy of urban streets. This composition should capture the essence of a bustling city environment, where the fusion of jazz elements creates a lively, urban atmosphere.\\ \hline
\end{longtable}
}

\section{Limitation Analysis of ComposerX}
This section is based on \citep{deng2024composerx}.

\subsection{Overview of ComposerX}

ComposerX is a symbolic music generation framework based on multi-agent collaboration \cite{deng2024composerx}. Its core lies in achieving multi-voice music composition through well-defined agent roles and structured communication processes. This architecture uses GPT-4-turbo as the base model, eliminating the need for additional training. By assigning roles and interaction rules, it activates the built-in music knowledge and reasoning capabilities of the LLM.

The system contains six role-playing agents, each with clear functions and collaborative cooperation:

\textbf{Group Leader}: Parses user input (such as music style, tonality, instrument requirements, etc.), decomposes tasks into sub-tasks such as melody creation, harmony design, and orchestration, and assigns them to corresponding agents.

\textbf{Melody Agent}: Generates a single-voice melody according to the group leader's instructions, follows music rules such as phrase division and rhythm coherence, and outputs in ABC notation format.

\textbf{Harmony Agent}: Adds harmony and counterpoint elements to the melody, ensures that the chord progression matches the melody's tonality, and enhances the musical layering.

\textbf{Instrument Agent}: Assigns instruments to the melody and harmony parts, combines the timbre characteristics of instruments with the music style, and optimizes the auditory effect.

\textbf{Reviewer Agent}: Evaluates the intermediate results from dimensions such as melody structure, harmony and counterpoint, rhythm complexity, orchestration rationality, and overall form, and provides modification feedback.

\textbf{Arrangement Agent}: Integrates the outputs of each agent, 
standardizes them into a unified ABC notation, and ensures that the final result can be interpreted by music software or performers.

The agents in the system collaborate through a closed-loop process of 'Initial Creation - Iterative Review - Final Arrangement':

\textbf{Initial Composition Round}: After the group leader assigns tasks, the melody, harmony, and instrument agents generate basic content in sequence.

\textbf{Iterative Review and Feedback Cycle}: The reviewer agent provides feedback on problems, and each creative agent corrects in the order of melody → harmony → instrument, repeating multiple rounds until the quality requirements are met.

\textbf{Final Arrangement and Notation}: The arrangement agent unifies the format and outputs a complete musical work.

The ComposerX system simulates the human collaborative creation scenario, reduces the cognitive load of a single model through division of labor, and at the same time reduces generation errors through multiple rounds of feedback, ultimately achieving polyphonic music composition that conforms to music theory norms and meets user needs.

\subsection{Limitations of ComposerX}

According to the data in the ComposerX paper, the current good case rate of its generated works is 18.4\%. That is to say, users may need to generate multiple times to get a satisfactory work. This greatly reduces the usability of the system. In addition, the author states in the paper that the ComposerX system still has many deficiencies in music composition:

\textbf{Lack of delicacy in musical expression}: Although it can interpret basic musical elements, it is difficult to create delicate works with the characteristics of human composers. There are deficiencies in emotional depth, dynamic contrast, and complex musical phrases, which are crucial for conveying profound musical narratives and experiences.

\textbf{Gap in the conversion from natural language to musical score}: Instructions and feedback from team leaders and review agents regarding delicate musical elements sometimes cannot be fully transformed into ABC notation by the music agent, indicating that there is a gap between concept understanding and actual musical score presentation when the system realizes complex musical ideas.

\textbf{Compliance issue of instrument range}: Occasionally, it generates notes that exceed the conventional range of a specific instrument. For example, for the double bass (with a range from C2 to F4), notes exceeding its upper limit may appear, which does not conform to the actual limitations of music performance.

\textbf{Difficulty in multi-voice alignment}: There are challenges in accurately aligning multiple musical voices. This is mainly due to the inherent limitations of text-based LLMs in generating polyphonic ABC notation. The linear input-output mode of text is difficult to adapt to the complexity of multiple voices or instruments that need to be coordinated in time in polyphonic music.

\textbf{Inadequate cadence resolution}: Some generated works lack a clear sense of ending, giving people a feeling of being unfinished or ending abruptly, affecting the audience's sense of closure and satisfaction. To some extent, this is because language models like GPT have difficulty understanding the concept of musical cadence, and their own characteristics make it difficult for them to handle this kind of musical ending problem.

\end{document}